\documentstyle[12pt,a4]{article}

\begin{document}

\author{F. Boudjema$^1$ and N. Dombey$^2$\thanks{%
SUSX-TH-95/47, Submitted to Physics Letters B} \\
$^1$Laboratoire de Physique Th\'eorique ENSLAPP\\
Chemin de Bellevue, B.P. 110, F-74941 Annecy-le-Vieux,\\
Cedex, France\\
$^2$School of Mathematical \& Physical Sciences, University of\\
Sussex, Falmer, Brighton, Sussex, BN1 9QH, UK}
\title{Threshold Behaviour in Gauge Boson Pair Production at LEP 2}
\maketitle

\begin{abstract}
\noindent We discuss the form of the amplitude for gauge boson pair
production at or near threshold.We show that in the case of W-pair
production at LEP2 near threshold only one anomalous electromagnetic
coupling can contribute. This anomalous coupling is CP violating and
contributes to the electric dipole moment of the $W$. Since this coupling is
likely to be small, it is important to look for $ZZ\gamma $ couplings in $%
Z\gamma $ production. These couplings are not suppressed at the $W$
threshold.

\newpage\
\end{abstract}

\section{Introduction}

CERN intends to increase the energy of its electron-positron collider LEP so
that $W$-pairs can be produced. The upgraded LEP 2 should demonstrate
directly the presence of the non-Abelian gauge coupling of the $Z$ to $%
W^{+}W^{-}$. It should also show the effects of the magnetic moment and the
quadrupole moment of the $W.\,$At tree-level, these quantities are predicted
by the standard electroweak model without any assumptions on its Higgs
content but have not yet been measured directly \footnote{%
Present limits from the Tevatron are too large to be meaningful} . The
energy of LEP 2 may eventually be increased to allow the production of $Z$%
-pairs in addition to $W$-pairs.

\smallskip\

Many authors including ourselves have considered alternatives to the
standard electroweak model in which additional self-couplings of the gauge
particles can occur \cite{Gounaris,HPZH,Mery,Anapoleus}. These new anomalous
couplings are often present in the standard model as small radiative
corrections, but in alternative models it is commonly speculated that they
may be large. For example, the $Z$ should couple to an electromagnetic field
through a $C$- and $P$-violating, but $CP$-conserving, anapole moment of
order $\alpha /\pi $ in the standard model\cite{anapsm}, but in a composite
model this anapole moment may be large, that is to say of order unity \cite
{Anapoleus}. Similarly, the $W$ has an electric dipole moment (edm) of order
less than 10$^{-38}e$ cm$^2$ \footnote{%
The electric dipole moment of the $W$ in the standard model is known to
vanish even at the two-loop order\cite{Khrip}. The bound we quote is from an
estimate given in reference \cite{Chang}.} in the minimal standard model,
but in alternative models it may be much larger and if so, this edm may even
be the source of the $CP$-violation observed in the $K^0-\bar K^0$ system.
Nevertheless, even if such anomalous couplings exist, they are not expected
to have any appreciable consequences in measurements at current LEP
energies. An important aim of LEP 2 is to search for the presence of these
anomalous couplings (and other variations from the predictions of the
standard model) in addition to its primary purpose of demonstrating the $WWZ$
non-Abelian coupling and the magnetic and quadrupole moments of the $W$.

\smallskip\

In this note we consider the general form of the three vector boson coupling
which gives in particular the electromagnetic structure of the $W$ and the $%
Z $. This allows us to formulate threshold theorems for all the gauge boson
pair production processes. In the next section we show that a threshold
theorem exists for $e^{+}e^{-}\rightarrow W^{+}W^{-}$ which picks out one
particular anomalous coupling. In section 3 we discuss $e^{+}e^{-}%
\rightarrow ZZ$ which is especially simple to analyse on account of the
Majorana (self-dual) nature of the $Z$. This property severely limits the
form of the $ZZ\gamma $ (or $ZZZ$) interaction and gives a very clean
threshold theorem. In section 4 we discuss $e^{+}e^{-}\rightarrow Z\gamma $,
which satisfies a threshold theorem but at LEP, not LEP 2 energies. It is
therefore especially important to study this process at LEP 2 since the
constraints of the threshold theorem no longer apply.

\section{Threshold Theorem for $W^{+}W^{-}$ Production}

The general form of the electromagnetic coupling of a spin-one particle of
mass $M$ can be written in the form \cite{HPZH}
\[
\Gamma ^{\alpha \beta \mu }\left( q,\bar q,k\right) =\tilde f_1\left( q-\bar
q\right) ^\mu g^{\alpha \beta }-\frac{\tilde f_2}{M^2}\left( q-\bar q\right)
^\mu k^\alpha k^\beta +\tilde f_3\left( k^\alpha g^{\mu \beta }-k^\beta
g^{\mu \alpha }\right)
\]

\begin{equation}  \label{nd}
\begin{array}{c}
+\,i\tilde f_4\left( k^2 (k^\alpha g^{\mu \beta }+ k^\beta g^{\mu \alpha
})-2k^\alpha k^\beta k^\mu \right) \\
\\
+i\tilde f_5\left( k^2\epsilon ^{\alpha \beta \mu \rho }\left( q-\bar
q\right) _\rho +2 k^\mu \epsilon ^{\alpha \beta \rho \sigma }q_\rho \bar
q_\sigma \right)
\end{array}
\end{equation}

\[
-\tilde f_6\epsilon ^{\mu \alpha \beta \rho }k_\rho -\frac{\tilde f_7}{M^2}%
\left( q-\bar q\right) ^\mu \epsilon ^{\alpha \beta \rho \sigma }k_\rho
\left( q-\bar q\right) _\sigma
\]

\noindent Here, $q$ and $\bar q$ are the momenta of the $W^{+}$ and $W^{-}$
and $k$ = $q$ + $\bar q$ is the virtual photon momentum. The $W$ is taken to
be on-shell so $q^2=\bar q^2=M^2$. Since the electromagnetic current is
conserved we have
\begin{equation}  \label{nd}
k_\mu \Gamma ^{\alpha \beta \mu }=0
\end{equation}

\noindent No symmetry other than electromagnetic current conservation and
Lorentz invariance is assumed in this decomposition.

\smallskip\

If we now consider the process $e^{+}$e$^{-}\rightarrow W^{+}W^{-}$ through
the virtual photon exchange term of Figure 1, we can ignore all terms in Eq.
(1) proportional to $k^\mu $ since the electromagnetic current of the
electron is also conserved. Hence we obtain the form of the coupling written
down by Hagiwara et al.\cite{HPZH}
\[
\Gamma _T^{\alpha \beta \mu }\left( q,\bar q,k\right) =f_1\left( q-\bar
q\right) ^\mu g^{\alpha \beta }-\frac{f_2}{M^2}\left( q-\bar q\right) ^\mu
k^\alpha k^\beta +f_3\left( k^\alpha g^{\mu \beta }-k^\beta g^{\mu \alpha
}\right)
\]

\begin{equation}  \label{nd}
+if_4\left( k^\alpha g^{\mu \beta }+k^\beta g^{\mu \alpha }\right)
+if_5\epsilon ^{\mu \alpha \beta \rho }\left( q-\bar q\right) _\rho
\end{equation}

\[
-f_6\epsilon ^{\mu \alpha \beta \rho }k_\rho -\frac{f_7}{M^2}\left( q-\bar
q\right) ^\mu \epsilon ^{\alpha \beta \rho \sigma }k_\rho \left( q-\bar
q\right) _\sigma
\]

The seven form factors $f_i$ are functions only of $k^2$:$f_1(0),\,\,f_2(0)$
and $f_3(0)$ determine the electromagnetic charge, magnetic moment and
quadrupole moment of the $W$ and conserve $C$, $P$ and $T$; $f_6$ and $f_7$
violate $P$ and $T$ but conserve $C$ and hence can lead to an electric
dipole moment of the $W$, while $f_4$ conserves $P$ but violates $C$ and $T$%
.The anapole $f_5$ conserves $T$ but violates $C$ and $P$. By comparing
Equations (1) and (3) we obtain the constraints $f_4(0)=f_5(0)=0$ imposed by
current conservation.

\smallskip\

In $W$-pair production at LEP 2 the amplitude to first order in $\alpha $ is
given by the Born approximation, that is the neutrino exchange term (Figure
2) which is non-zero at threshold, the virtual photon exchange term of
Figure 1, together with a corresponding term in which the virtual photon is
replaced by a virtual $Z$. The We$\nu $ coupling is known from $\mu $-decay
and $\beta $-decay.

\smallskip\

The description of virtual $Z$-exchange is almost identical to that of
virtual photon exchange. We now take the $ZWW$ coupling (or weak vertex
function of the $W$) to be given by $G^{\alpha \beta \mu }$ instead of $%
\Gamma ^{\alpha \beta \mu }$ and we can break up $G^{\alpha \beta \mu }$
into its transverse and longitudinal parts

\begin{equation}  \label{nd}
G^{\alpha \beta \mu }=G_T^{\alpha \beta \mu }+G_L^{\alpha \beta \mu }
\end{equation}

\noindent where $G_L^{\alpha \beta \mu }$ includes all terms proportional to
$k^\mu$. \ \noindent Provided that we take $m_e=0$, the weak neutral current
of the electron- positron system which couples to the $Z$ is conserved. This
ensures that the longitudinal part, $G_L^{\alpha \beta \mu }$, 
does not contribute to the amplitude. So we can take the $ZWW$ coupling to
be given by $G_T^{\alpha \beta \mu }$ and this can be decomposed in exactly
the same form as the electromagnetic vertex $\Gamma_T ^{\alpha \beta \mu }$
in Eq. 1 provided we simply replace the form factors $f_i$ by the
corresponding form factors $f_i^Z$. We can also neglect Higgs exchange in
the $m_e=0$ limit.

\smallskip\

At threshold $q=\bar q$ in the centre of mass frame and hence $f_1,$ $f_2$
and $f_7$ do not contribute to the virtual photon exchange diagram.
Furthermore in this frame $k$, $q$ and $\bar q$ are purely timelike while
the polarization vectors $\epsilon $ and $\bar \epsilon $ of $W^{+}$ and $%
W^{-}$ are purely spacelike. Hence when contracted with $\epsilon ^\alpha $,
$\bar \epsilon ^\beta $, $f_3$ and $f_4$ are coefficients of $k.\epsilon $
and $k.\bar \epsilon $ both of which vanish. So only the CP-violating term $%
f_6$ persists at threshold. For $Z$-exchange a similar argument gives $f_6^Z$
as the sole contribution.

\smallskip\

So we end up with the threshold theorem for $W$-pair production at threshold

\begin{equation}  \label{nd}
\frac{{\rm d}\sigma \left( e^{+}e^{-}\rightarrow W^{+}W^{-}\right)} {{\rm d}%
\Omega} = \frac{1}{s} \left(\frac{M_W^2 G_\mu}{\pi \sqrt{2}}\right)^2 \beta
\left[ 1+ 4\beta \cos\theta \frac{3 c_W^2 -1}{4 c_W^2-1} + 2 s_W^4 F_6^2 +
{\cal O}(\beta^2) \right]
\end{equation}

where we define
\begin{eqnarray}
F_6^2=f_6^2+ \frac{{{\hat f}_{6,Z}^2}}{2} (1+(1-\frac{ 1}{2s_W^2})^2) -2f_6{%
\hat f}_{6,Z}(1-\frac {1}{4s_W^2}) \\
{\hat f}_{6,Z}=\frac{ s}{s-M_Z^2}f_6^Z\;\; \;\; ; \; \;\;\; c_W^2=1-s_W^2=%
\frac{M_W^2}{M_Z^2}  \nonumber
\end{eqnarray}

Note that the interference term between the CP violating $Z$ and photon
exchange diagrams is suppressed. As expected the leading order terms in $%
\beta $ (the neutrino exchange t-channel and CP violating s-channel) give an
isotropic angular distribution. The next order term in $\beta $ is solely
from the t-channel and leads to an asymmetry that vanishes upon integration.

\smallskip\

The corollary of this result is that if the electric dipole moment $f_6$
(and any associated weak dipole moment $f_6^Z$) of $W$ is too small to be
measurable, then the standard model dominates in $W$-pair production near
threshold as the other $WW\gamma $ and $WWZ$ couplings all enter into the
cross-section with factors of at least $\beta $, where $\beta $ is the
velocity of the $W$. This result just follows from kinematics together with
electromagnetic (and weak) current conservation, just as it does in classic
low energy theorems.\cite{low} It requires no assumptions on the underlying
gauge structure of the theory. Furthermore as the We$\nu $ coupling is taken
from its value as observed in $\beta $-decay the Born approximation includes
the radiative corrections associated with the running of the fine structure
constant.

\smallskip\

It can be expected that an electric dipole moment $f_6$ (and any associated
weak dipole moment $f_6^Z$) of the $W$ would contribute to the neutron
electric dipole moment. The experimental limits on the neutron edm allow
bounds to be set on the $W$ edm. These bounds, however, are not rigorous
\cite{indcp} and LEP 2 can be expected to provide a more substantive result.

\smallskip\

In this discussion we have not considered the radiative corrections to the
process, in particular the initial state radiation effects which give the
dominant contribution (see the treatment by Bardin et al \cite{Bardin} and
Berends, Pittau and Kleiss \cite{Pittau}). Nor have we considered the finite
width effects which can be substantial in the threshold region. These may be
treated according to the prescription of Fadin, Khoze and Martin.\cite{Fadin}

\section{Threshold Theorem for $e^{+}e^{-}\rightarrow ZZ$}

Extensive studies have been devoted to anomalous $WW\gamma $ and $WWZ$
couplings but rather less attention has been focussed on the anomalous $%
ZZ\gamma $ and $ZZZ$ couplings. There may be three reasons for this: (a)
within standard electroweak theory the $WW$ cross section is an order of
magnitude larger than the $ZZ$ cross section; (b) at LEP 2 there is not
expected to be sufficient energy initially to produce $Z$-pairs and (c) $WWZ$
and $WW\gamma $ couplings appear at tree level in standard electroweak
theory $ZZ\gamma $ and $ZZZ$ couplings are absent.

\smallskip\

Nevertheless, from the purely phenomenological point of view that we have
adopted here the $ZZ\gamma $ and $ZZZ$ coupling are particularly easy to
analyse. This is because the $Z$ is a Majorana particle\cite{Hamz} (i.e. it
is its own antiparticle) and hence, using the same notation that we used in
the previous section, the $ZZ\gamma $ on-shell coupling must be symmetric
under interchange of the two $Z$s; that is to say that

\begin{equation}  \label{nd}
\Gamma ^{\alpha \beta \mu }(q,\bar q,k^2)=\;\Gamma ^{\beta \alpha \mu }(\bar
q,q,k^2).
\end{equation}

\noindent By inspection it is easily seen that only the anapole couplings $%
\tilde{f}_4(k^2)$ and $\tilde{f}_5(k^2)$ satisfy this criterion. Imposing CP
conservation only $\tilde{f}_5(k^2)$ survives. Likewise, for $ZZZ$ coupling,
the corresponding weak anapole moment $\tilde{f}_5^Z$ is the only term,
provided that the first two $Z$s are on-shell and the third off-shell.

\smallskip\

So the threshold theorem for $Z$-pair production follows simply: the
cross-section at threshold is given by the standard model since the $Z$
cannot possess an electric dipole moment. At higher energies the anapole
term which is pure $P$ wave with a (1+$\cos ^2\theta $) angular distribution
should play a role. Note that whereas in $W$-pair production above threshold
up to six anomalous couplings may be present outside the threshold region,
in $Z$-pair production there is only the anapole. The anapole couples a
longitudinal $Z$ to a transverse $Z$, so the signal for the presence of the
anapole would arise in the $\sigma _{TL}$ cross-section. The unique CP
conserving $ZZ\gamma $ operator which contributes to $ZZ$ production is
given by
\begin{equation}
{\cal L}_A\;=\;\frac{e\kappa _A}{\Lambda _A^2}\;\;F_{\mu \nu }\;\;\partial
^\mu \;(\tilde Z^{\alpha \nu }Z_\alpha )
\end{equation}

\noindent
In momentum space this operator is proportional to the invariant mass $k^2$
of the photon (see Eq.~1) which means that in $e^{+}e^{-}$ processes this
leads to a contact interaction\cite{Hamz}. The $ZZZ$ vertex on account of
Bose symmetry is of the same Lorentz structure as the anapole with the $%
(k^2) $ term replaced by $k^2-M_Z^2$: Again it also leads to a contact
interaction. For these reasons unless the initial beams are polarised we
cannot distinguish between a $ZZZ$ and $ZZ\gamma $ anomalous coupling in $ZZ$
production. For illustration we show the contribution of the electromagnetic
operator to the differential $ZZ$ cross section

\begin{equation}  \label{nd}
\frac{d\sigma _{TL}}{d\cos \theta }=\frac{{\pi \alpha ^2\left( M_Z^2\right) }%
}{M_Z^2}\beta \left\{ G_{ZZ}\frac{4\cos ^2\theta +\left( 2-\gamma \sin
^2\theta \right) ^2}{\left( 4+\gamma ^2\beta ^2\sin ^2\theta \right) }%
+\left( \frac{k_AM_Z^2}{4\Lambda_A ^2}\right) ^2\gamma ^2\beta ^4(1+\cos
^2\theta )\right\}
\end{equation}

\noindent where $\beta ^2=1-4M_{Z\,}^2/s$, $\gamma =s/M_Z^2$ and $%
G_{ZZ}=G_{Z\gamma }^2+4X_{Z\gamma }^2.$ The quantities
\begin{equation}  \label{nd}
G_{Z\gamma }=\frac{\left( 1-4s_W^2\right) ^2+1}{16s_W^2c_W^2}
\end{equation}

\noindent and
\begin{equation}  \label{nd}
X_{Z\gamma }=-\frac{\left( 1-4s_W^2\right) }{16s_W^2c_W^2}
\end{equation}

\noindent will occur in $e^{+}e^{-}\rightarrow Z\gamma $.

\smallskip\ \

\noindent Note that the anomalous term is strongly suppressed by a function $%
\beta ^4$ compared to the standard model terms where $\beta $ is the $Z$
velocity. So energies well above the planned LEP 2 energies would be
necessary to investigate this coupling. Nevertheless it is perhaps worth
noting that at a future electron-positron collider of energy 500GeV or more,
it should be relatively easy to disentangle the standard model terms from
the anomalous anapole term. This would be accomplished by a cut in
scattering angle about the forward direction which would severely restrict
those events arisng from the standard model while hardly affecting the
anomalous terms.

\section{Threshold Theorem for $e^{+}e^{-}\rightarrow Z\gamma $}

We showed some years ago \cite{theozg} (see also \cite{Berends}) that the
threshold amplitude for $e^{+}e^{-}\rightarrow Z\gamma $ was determined by
the $Ze^{+}e^{-}$ coupling constant including one-loop radiative
corrections. This is simply a statement that the threshold energy for this
process is the $Z$ mass and therefore soft-photons have to be treated
according to quantum electrodynamics in the usual way. So near threshold,
the standard model term for this process which correctly contains the
threshold amplitude must dominate over all anomalous terms.

\smallskip\

The $Z$-exchange diagram for the process $e^{+}e^{-}\rightarrow Z\gamma $
where one $Z$ is on-shell and one is off-shell and the photon is real,
involves a coupling of electric dipole transition type that vanishes when
both $Z$'s become real. The CP conserving operator that parametrizes this
transition can be written as

\begin{equation}
{\cal L}_E\;=\; \frac{e \kappa_E}{\Lambda_{E}^{2}} \;\; Z_{\mu \nu} \;\;
\partial^{\mu}(\tilde{F}^{\alpha \nu} Z_{\alpha} ) \;\;\;\;\;\;\; \tilde{F}%
^{\mu \nu}=\frac{1}{2} \epsilon^{\mu \nu \lambda \rho} F_{\lambda \rho}
\end{equation}

\noindent
where $F^{\mu \nu }$ is the electromagnetic tensor, $\Lambda _E$ is the
scale associated to this operator and $\kappa _E$ characterises the strength
of the (EDT) coupling. It is useful to introduce the dimensionless parameter
$k_E=\kappa _EM_Z^2/\Lambda _E^2$\\One can already set a bound on this
operator from the LEP data since it contributes to the radiative decay $%
Z\rightarrow f\overline{f}\gamma $. We have:

\begin{equation}
Br^{(EDT)}\;=\; \frac{\Gamma^{(EDT)} (Z \rightarrow f \bar{f} \gamma)}{%
\Gamma_{SM} (Z \rightarrow f \bar{f}) } \;=\; \frac{\alpha}{80 \pi} \;\;
k_E^2
\end{equation}

\noindent
The best limit is from the search of single (energetic) isolated photons
associated with $\nu \bar \nu \gamma $ production. The L3 Collaboration\cite
{L3} gives the limit $|k_E|\leq 0.80\;@95\%$ C.L. In fact this limit
corresponds to all events collected for centre-of-mass energies $88.56<\sqrt{%
s}<93.75GeV$. This is not as good a limit as one might hope to get due to
the P-wave nature of the transition which makes the photon prefer to have as
large an energy as possible, but around the $Z$ peak this is offest by the
fact that this configuration forces the intermediate Z to be far from
resonance.\\

Therefore, in order to look for this anomalous coupling, it is necessary as
in $e^{+}e^{-}\rightarrow ZZ$ to be well above threshold. It turns out that
while the cross section for the production of a transverse $Z$ in the
standard model term is strongly peaked in the forward/backward directions,
the cross section for the production of a longitudinal $Z$ has an isotropic
angular distribution. So including the anomalous transition it is easy to
show that
\begin{equation}  \label{nd}
\frac{d\sigma _L}{d\cos \theta }=\frac{4\pi \alpha \left( M_Z^2\right) }%
sG_{Z\gamma }\frac 1{\beta \gamma }\left\{ \left[ 1 +k_{E}\gamma^2\beta ^2%
\frac{X_{Z\gamma }}{2G_{Z\gamma }}+\left( \frac{k_E}8\right) ^2\gamma
^4\beta ^4\right] +\left( \frac{k_E}8\right) ^2\beta ^4\gamma ^4\cos
^2\theta \right\}
\end{equation}

\ \noindent where now $\beta =1-M_Z^2/s,\,\gamma =s/M_Z^2$, and $G_{Z\gamma}$
$X_{Z\gamma }$ were defined above.

\noindent Note that again the anomalous term is suppressed near threshold by
a factor $\beta ^4$.

\smallskip\

The importance of $e^{+}e^{-}\rightarrow Z\gamma $ compared to the first two
processes is that at foreseeable LEP 2 energies, this process is always well
above threshold while the first two processes we have discussed are likely
to be still in the threshold region. Hence while it will be difficult to
detect any anomalous couplings in e$^{+}e^{-}\rightarrow WW$ at LEP 2 if
CP-violating couplings are small and if LEP 2 is operated at or very close
to threshold, and essentially impossible to detect anomalous couplings of
the $Z$ in $e^{+}e^{-}\rightarrow ZZ$; there should be no difficulty in
searching for anomalous couplings of the $Z$ in $e^{+}e^{-}\rightarrow
Z\gamma $.

\section{Conclusions}

We have shown that if CP is nearly conserved, then it will be difficult to
measure deviations from the standard model in $W$-pair production at LEP2.
The likelihood of a successful measurement of any deviation increases
substantially with the energy of the machine. It is therefore essential to
work at the highest energies possible. \smallskip\ In some ways this result
is reminiscent of that of de Rujula et al.\cite{Ruj}, who state that the
standard model must be satisfied in the LEP2 energy region. Our result
differs from theirs, however, in several ways: first their result makes no
distinction between threshold and non-threshold energies; second our
emphasis is on CP-conservation and not SU(2)xU(1) local gauge invariance,
and third if CP is badly broken in the gauge boson sector then a clear
discrepancy from the standard model will be observed. \smallskip\ Finally we
stress again the importance of looking for at $Z\gamma $ production at LEP2
since this process is well above threshold.

\

\section{Figure Captions}

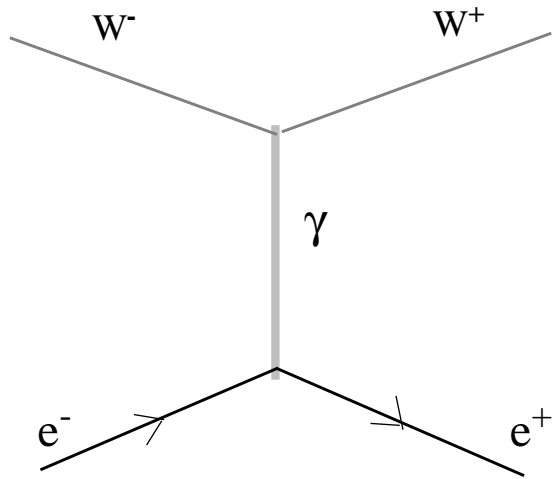
Figure 1: $e^{+}e^{-}\rightarrow W^{+}W^{-}$ via a virtual photon.\\
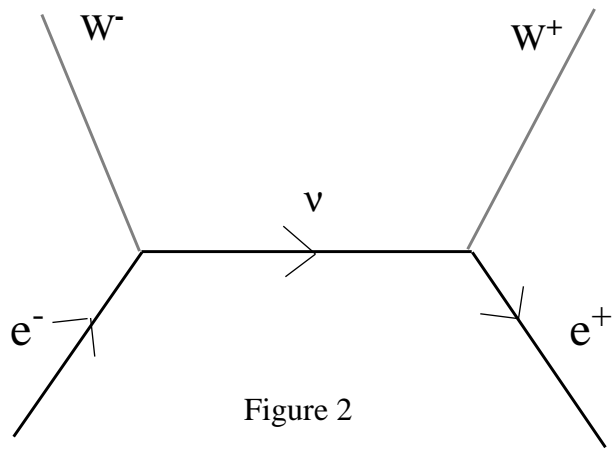
Figure
2: $e^{+}e^{-}\rightarrow W^{+}W^{-}$ by means of neutrino exchange.

\end{document}